\documentstyle[twoside,fleqn,npb,epsfig]{article}
\newcommand{\AmS}{{\protect\the\textfont2
  A\kern-.1667em\lower.5ex\hbox{M}\kern-.125emS}}

\title{Mass effects in the polarized virtual photon structure
 \thanks{Talk presented by T. Uematsu at Loops and Legs 2004, 
Zinnowitz, Germany, April 25-30, 2004.\ YNU-HEPTh-04-101; KUNS-1922.
}}
\author{Ken Sasaki\address{Department of Physics, Faculty of Engineering,
        Yokohama National University, \\ Yokohama 240-8501, Japan}
        and Tsuneo Uematsu\address{Department of Physics, Graduate School of Science,         Kyoto University, \\
        Kyoto 606-8501, Japan}%
}

\begin{document}

\begin{abstract}
We discuss target mass effects in the polarized virtual photon structure 
functions $g_1^\gamma (x,Q^2,P^2)$, $g_2^\gamma (x,Q^2,P^2)$
for the kinematic region $\Lambda^2\ll P^2 \ll Q^2$, where 
$-Q^2 (-P^2)$ is the mass squared of the probe (target) photon.
We obtain the expressions for the structure functions
in closed form by inverting the
Nachtmann moments for the twist-2 and twist-3 operators. Numerical
analysis shows that  target mass effects 
appear at large $x$ and become sizable near 
the maximal value of  $x$, 
as the ratio  $P^2/Q^2$ increases.
Target mass effects for the sum rules 
of $g_1^\gamma$ and $g_2^\gamma$ 
are also discussed.
\end{abstract}

\maketitle

\section{INTRODUCTION}

We would like to talk about the virtual photon structure,
especially on the mass effects in the polarized photon structure 
functions. 
In the last several years, there has been much interest in the
spin-dependent photon structure functions which can be studied
in the polarized version of the $ep$ collider or more directly in
the polarized $e^+e^-$ collision in the future linear 
collider (Figure.1).

\vspace{-1cm}

\begin{figure}[htbp]
\begin{center}
\epsfxsize=5cm
\hspace*{0cm}
\ \epsfbox{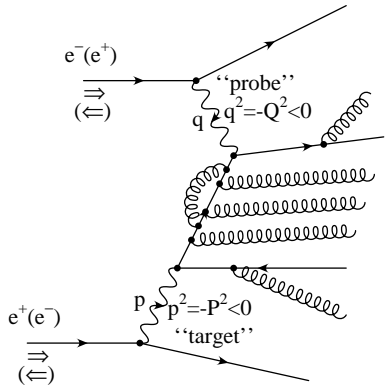}
\vspace{-1cm}
\caption{Two-photon process in polarized $e^+e^-$ collision}
\end{center}
\end{figure}
\vspace*{-1.3cm}

Now let us consider the two-photon process in the polarized 
electron-positron collision where both of the photons are off-shell. 
We particularly
consider the case where the mass square of the \lq\lq probe\rq\rq photon,
$Q^2$, is much bigger than that of the \lq\lq target\rq\rq photon, $P^2$,
which is in turn much bigger than the square of the QCD scale,
$\Lambda^2$. 
The advantage for studying the virtual photon target in 
this kinematical region is that we can calculate whole structure functions
$g_1^\gamma$ and $g_2^\gamma$ in perturbative QCD
up to next-leading-order (NLO), in contrast to the real photon target where
there remain uncalculable non-perturbative pieces. 
Note that the 1st moment of $g_1^\gamma$ is
related to the axial anomaly just like the nucleon case, while the
second structure function $g_2^\gamma$ only exists for the virtual photon
target ($P^2\neq 0$).

Now the possible mass effects are twofold; target-mass effects and
quark-mass effects. Here in this talk we discuss the former, which 
appear as power-corrections in $P^2/Q^2$.
For the real photon target ($P^2=0$),  there is no need to consider target
mass corrections. But when the target becomes off-shell and 
for relatively 
low values of $Q^2$, contributions suppressed by powers of 
$\frac{P^2}{Q^2}$ may become important. Then we need to
take into account these target mass contributions just like the case of 
the nucleon structure functions. The consideration of target mass effects 
(TME) is important by another reason. 
For the virtual photon target, the maximal value of the Bjorken variable $x$ 
is not 1 but 
\begin{center}
\begin{equation}
x_{\rm max}=\frac{1}{1+\frac{P^2}{Q^2}}~, \label{xmax}
\end{equation}
\end{center}
due to the constraint $(p+q)^2 \ge 0$, which is contrasted with  
the nucleon case where $ x_{\rm max}= 1$.
The structure functions should 
vanish at $x=x_{\rm max}$. However, the NLO QCD result \cite{SU} for   
$g_1^\gamma(x, Q^2, P^2)$ (See also the second paper of \cite{GRS})
shows that the predicted graph does not vanish but
remains finite  at $x=x_{\rm max}$. 
In this talk we discuss the TME for $g_1^\gamma(x, Q^2, P^2)$ and
$g_2^\gamma(x, Q^2, P^2)$ in the framework of operator product expansion (OPE)
supplemented by the renormalization group (RG) method \cite{BSU03}.

\section{STRUCTURE FUNCTIONS}

Let us consider the structure tensor $W_{\mu\nu\rho\tau}(p,q)$ which
is the absorptive part of the forward scattering amplitude 
$T_{\mu\nu\rho\tau}(p,q)$ (Figure 2):
\begin{equation}
W_{\mu\nu\rho\tau}(p,q)=\frac{1}{\pi}{\rm Im}T_{\mu\nu\rho\tau}(p,q)~.
\end{equation}
for the target photon with mass squared $p^2=-P^2$ probed by the photon 
with $q^2=-Q^2$.
\vspace{-1cm}
\begin{figure}[htbp]
\begin{center}
\epsfxsize=4cm
\hspace*{0cm}
\ \epsfbox{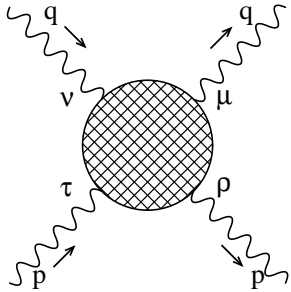}
\vspace{-1cm}
\caption{Virtual photon-photon scattering}
\end{center}
\end{figure}
\vspace{-1cm}

The antisymmetric part $W_{\mu\nu\rho\tau}^{A}$ under 
$\mu\leftrightarrow\nu$ and $\rho\leftrightarrow\tau$,  can be decomposed as
\begin{eqnarray}
&&\hspace{-.7cm}W_{\mu\nu\rho\tau}^{A}
=\epsilon_{\mu\nu\lambda\sigma}q^\lambda
{\epsilon_{\rho\tau}}^{\sigma\beta}p_\beta \frac{1}{p\cdot
q}g_1^\gamma\nonumber\\
&&\hspace{-.7cm}+
\epsilon_{\mu\nu\lambda\sigma}q^\lambda
(p\cdot q\ {\epsilon_{\rho\tau}}^{\sigma\beta}p_\beta-\epsilon_{\rho\tau
\alpha\beta}p^\beta p^\sigma q^\alpha )\frac{1}{(p\cdot q)^2}g_2^\gamma\nonumber\\
\end{eqnarray}
which gives two spin-dependent structure functions, $g_1^\gamma(x,Q^2,P^2)$ 
and $g_2^\gamma(x,Q^2,P^2)$. When the target is real photon ($P^2= 0$), $g_2^\gamma$
is identically  zero, and there exists only one spin structure function,
$g_1^\gamma(x,Q^2)$.  
While, for the off-shell or virtual photon ($P^2
\neq 0$) target,  we have two 
spin-dependent structure functions $g_1^\gamma$ and $g_2^\gamma$.

The deep inelastic photon-photon scattering (Figure 2.) 
amplitude relevant for the
polarized structure functions 
is given by
\begin{eqnarray}
&&\hspace{-0.5cm}T^A_{\mu\nu\rho\tau}=i\int d^4x e^{iq\cdot x}
\nonumber\\
&&\hspace{-0.5cm}\times
\langle 0|T(A_\rho(-p)(J_\mu(x)J_\nu(0))^AA_\tau(p))|0\rangle_{\rm Amp}.
\label{Amplitude}
\end{eqnarray}
For the product of the two electromagnetic currents we apply the OPE
and obtain for the $\mu$-$\nu$ antisymmetric part
\begin{eqnarray}
&&\hspace{-0.5cm}i\int d^4x e^{iq\cdot x}T(J_\mu(x)J_\nu(0))^A \nonumber\\
&&\hspace{-0.5cm}=-i\epsilon_{\mu\nu\lambda\sigma}q^\lambda
\sum_{n=1,3,\cdots}\left(\frac{2}{Q^2}\right)^n
q_{\mu_1}\cdots q_{\mu_{n-1}}\nonumber\\
&&\hspace{-0.5cm}
\times
\left\{
\sum_i E_{(2)i}^n R_{(2)i}^{\sigma\mu_1\cdots\mu_{n-1}}
+\sum_i E_{(3)i}^n R_{(3)i}^{\sigma\mu_1\cdots\mu_{n-1}}
\right\}\label{CurrentProdFourier}\nonumber\\
\end{eqnarray}
where $R^n_{(2)i}$ and $R^n_{(3)i}$ are the twist-2 and twist-3 operators,
respectively, and  are both traceless, and $E_{(2)i}^n$ and $E_{(3)i}^n$ are 
corresponding coefficient functions.
The twist-2 operators $R^n_{(2)i}$ have totally symmetric Lorentz indices
$\sigma\mu_1\cdots\mu_{n-1}$, while the indices of twist-3 operators
$R^n_{(3)i}$  are totally symmetric among $\mu_1\cdots\mu_{n-1}$ but
antisymmetric under $\sigma \leftrightarrow \mu_i$. 

For the photon target we evaluate ``{\it matrix elements}" of the traceless 
operators $R^n_{(2)i}$ and $R^n_{(3)i}$  sandwiched by 
two photon states with momentum $p$,  which are written in the following forms:
\begin{eqnarray}
&&
\langle 0\vert T(A_{\rho}(-p)R_{(2)i}^{\sigma\mu_1\cdots\mu_{n-1}}A_{\tau}(p))
\vert 0\rangle_{\rm Amp}\nonumber\\
&&\hspace{0.5cm}=-ia_{(2)i}^{\gamma,n}
M_{(2)\rho\tau}^{\sigma\mu_1\cdots\mu_{n-1}}~, \label{matTwist2}\\
&&
\langle 0\vert T(A_{\rho}(-p)R_{(3)i}^{\sigma\mu_1\cdots\mu_{n-1}}A_{\tau}(p))
\vert 0\rangle_{\rm Amp}\nonumber\\
&&\hspace{0.5cm}=-ia_{(3)i}^{\gamma,n}
M_{(3)\rho\tau}^{[\sigma,\{\mu_1]\cdots\mu_{n-1}\}}~,\label{matTwist3}
\end{eqnarray}
where the subscript \lq Amp\rq\ stands for the amputation of  external
photon lines, $a_{(2)i}^{\gamma,n}$ and $a_{(3)i}^{\gamma,n}$  are reduced 
photon matrix elements. The traceless
tensors $M_{(2)\rho\tau}^{\sigma\mu_1\cdots\mu_{n-1}}$ and
$M_{(3)\rho\tau}^{[\sigma,\{\mu_1]\cdots\mu_{n-1}\}}$ 
satisfy the traceless conditions for $k=2,3$:
\begin{equation}
  g_{\sigma\mu_i}M_{(k)\rho\tau}^{\sigma\mu_1\cdots\mu_{n-1}}=0,\
  g_{\mu_i\mu_j}M_{(k)\rho\tau}^{\sigma\mu_1\cdots\mu_{n-1}}=0.
\end{equation}
Taking the \lq\lq {\it matrix elements}\rq\rq\ of
(\ref{CurrentProdFourier}) with the virtual photon states,
we obtain the deep-inelastic photon-photon forward 
scattering amplitude.

The basic idea for treating  target mass corrections exactly is 
to take account of trace terms in the traceless tensors properly
\cite{NACHT,GP}.
We evaluate the contraction between $q_{\mu_1}\cdots q_{\mu_{n-1}}$ and the traceless tensors without neglecting any of the trace terms.
The results are expressed in terms of Gegenbauer polynomials.

\section{NACHTMANN MOMENTS}

Now we follow the same procedures as were taken by Wandzura~\cite{WAND} and
in Ref.~\cite{MU}  for the polarized nucleon case, and we obtain the
analytic expression of  the Nachtmann moments \cite{NACHT}
for the twist-2 and twist-3 operators with definite spin $n$. 
By writing down the dispersion relations for the amplitudes and
with the use of orthogonality relations as well as
an integration formula for 
Gegenbauer polynomials $C_{n}^{(\nu)}(\eta)$, we project out 
$\sum_i a_{(2)i}^{\gamma,n}E_{(2)i}^n$ and 
$\sum_i a_{(3)i}^{\gamma,n}E_{(3)i}^n$ with definite spin $n$, 
which still include the infinite series in powers of $P^2/Q^2$. 
We then sum up those infinite series and express them in compact 
analytic forms \cite{MU}. Then we obtain
the Nachtmann moments \cite{BSU03}:
\begin{eqnarray}
&&\hspace{-0.5cm}
M_2^n\equiv \sum_i a_{(2)i}^{\gamma,n}E_{(2) i}^n(Q^2,P^2,g)\nonumber\\
&&\hspace{-0.5cm}=\int_0^{x_{\rm max}}\frac{dx}{x^2}{\xi}^{n+1}\left[
\left\{
\frac{x}{\xi}+\frac{n^2}{(n+2)^2}\frac{P^2 x\xi}{Q^2}
\right\}\right.   \nonumber\\
&&\hspace{-0.5cm}\left. \times
g_1^\gamma(x,Q^2,P^2)+\frac{4n}{n+2}\frac{P^2x^2}{Q^2}g_2^\gamma(x,Q^2,P^2)
\right] \nonumber\\
&&\hspace{-0.5cm}\quad (n=1,3,\cdots),  \label{Nacht2}\\
&&\hspace{-0.5cm}M_3^n \equiv \sum_i a_{(3)i}^{\gamma,n}E_{(3) i}^n(Q^2,P^2,g)\nonumber\\
&&\hspace{-0.5cm}=\int_0^{x_{\rm max}}\frac{dx}{x^2}{\xi}^{n+1}\left[
\frac{x}{\xi}g_1^\gamma(x,Q^2,P^2)\right.\nonumber\\
&&\hspace{-0.5cm}\left.+\left\{\frac{n}{n-1}\frac{x^2}{\xi^2}+
\frac{n}{n+1}\frac{P^2x^2}{Q^2} \right\}g_2^\gamma(x,Q^2,P^2) \right] 
\nonumber\\ 
&&\quad (n=3,5, \cdots),
\end{eqnarray}
where $x=Q^2/(2p\cdot q)$ and $\xi$, the so-called $\xi$-scaling
variable \cite{RGP}, is given by
\begin{equation}
\xi=\frac{2x}{1+\sqrt{1-\frac{4P^2x^2}{Q^2}}}~. \label{xi}
\end{equation}
The allowed range of $\xi$ is  $0\le \xi\le1$,
and $\xi(x_{\rm max})=1$. 
In the nucleon case, 
the constraint $(p+q)^2\geq M^2$ gives $x_{\rm max}=1$ and 
$\xi(x=1)<1$, leading to the problem of non-vanishing structure function at $x=1$.
The resolution to this problem was argued in refs. \cite{RGP,PR,BK},
by considering the dynamical higher-twist effects.
$M_2^n$ and $M_3^n$ 
are perturbatively calculable.
In fact, the perturbative QCD calculation of $M_2^n$ has been done in
LO~\cite{KS} and  in NLO~\cite{SU,GRS}, while the QCD analysis of $M_3^n$ has 
been carried out in LO for the flavor non-singlet part in the limit of large 
$N_c$~\cite{BSU}.


Once the moments $M_2^n$ and $M_3^n$ are known, we can derive 
$g_1^\gamma(x,Q^2,P^2)$ and $g_2^\gamma(x,Q^2,P^2)$ as functions of $x$ by
inverting  $M_2^n$ and $M_3^n$ as follows:
\hspace{-0.5cm}
\begin{eqnarray}
&&\hspace{-0.5cm}g_1^\gamma(x,Q^2,P^2)\nonumber\\
&&\hspace{-0.5cm}=4\kappa\xi^2~
\frac{(1+\kappa\xi^2)^3}{(1-\kappa\xi^2)^5}\left\{
1+\frac{2\kappa\xi^2}{(1+\kappa\xi^2)^2}  \right\}H_a(\xi)\nonumber\\
&&\hspace{-0.5cm}-4\kappa\xi^2~
\frac{(1+\kappa\xi^2)^2}{(1-\kappa\xi^2)^4}\left\{
1+\frac{1}{1+\kappa\xi^2}  \right\}G_a(\xi)\nonumber\\
&&\hspace{-0.5cm}
+\xi~\frac{(1+\kappa\xi^2)^2}{(1-\kappa\xi^2)^3}F_a(\xi)\nonumber\\
&&\hspace{-0.5cm}-8\kappa\xi^2~
\frac{(1+\kappa\xi^2)^3}{(1-\kappa\xi^2)^5}\left\{
1+\frac{2\kappa\xi^2}{(1+\kappa\xi^2)^2}  \right\}H_d(\xi)\nonumber\\
&&\hspace{-0.5cm}+12\kappa\xi^2~
\frac{(1+\kappa\xi^2)^2}{(1-\kappa\xi^2)^4}G_d(\xi)
-4\kappa\xi^3~\frac{1+\kappa\xi^2}{(1-\kappa\xi^2)^3}F_d(\xi)\nonumber\\
&&\label{TargetMass2}\\
&&\hspace{-0.5cm}g_2^\gamma(x,Q^2,P^2)\nonumber\\
&&\hspace{-0.5cm}=-6\kappa\xi^2~
\frac{(1+\kappa\xi^2)^3}{(1-\kappa\xi^2)^5}H_a(\xi)\nonumber\\
&&\hspace{-0.5cm}+\frac{(1+\kappa\xi^2)^3}{(1-\kappa\xi^2)^4}\left\{
1+\frac{4\kappa\xi^2}{1+\kappa\xi^2}  \right\}G_a(\xi)\nonumber\\
&&\hspace{-0.5cm}-\xi~\frac{(1+\kappa\xi^2)^2}{(1-\kappa\xi^2)^3}F_a(\xi)
+12\kappa\xi^2~
\frac{(1+\kappa\xi^2)^3}{(1-\kappa\xi^2)^5}H_d(\xi)
\nonumber\\
&&\hspace{-0.5cm}
-\frac{(1+\kappa\xi^2)^4}{(1-\kappa\xi^2)^4}
\left\{1+\frac{8\kappa\xi^2}{(1+\kappa\xi^2)^2}  \right\}
G_d(\xi)\nonumber\\
&&\hspace{-0.5cm}+\xi\frac{(1+\kappa\xi^2)^3}{(1-\kappa\xi^2)^3}F_d(\xi)~. \label{TargetMass3}
\end{eqnarray}
where $\kappa=P^2/Q^2$ and we have 
introduced the following functions first discussed in ref. \cite{PR}:
\vspace*{-1.0cm}
\begin{eqnarray}
H_{a,d}(\xi)&=&\frac{1}{2\pi i}\int_{c-i\infty}^{c+i\infty}
dn \,\xi^{-n} \frac{M_{2,3}^n}{n^2} , \quad \nonumber\\
G_{a,d}(\xi)&=&\frac{1}{2\pi i}\int_{c-i\infty}^{c+i\infty}
dn \,\xi^{-n}\frac{M_{2,3}^n}{n}, \quad  \nonumber\\
\xi F_{a,d}(\xi)&=&\frac{1}{2\pi i}\int_{c-i\infty}^{c+i\infty}
dn \,\xi^{-n}{M_{2,3}^n}~. 
\end{eqnarray}
\vspace{-3.5cm}
\begin{figure}[h] 
\epsfig{file=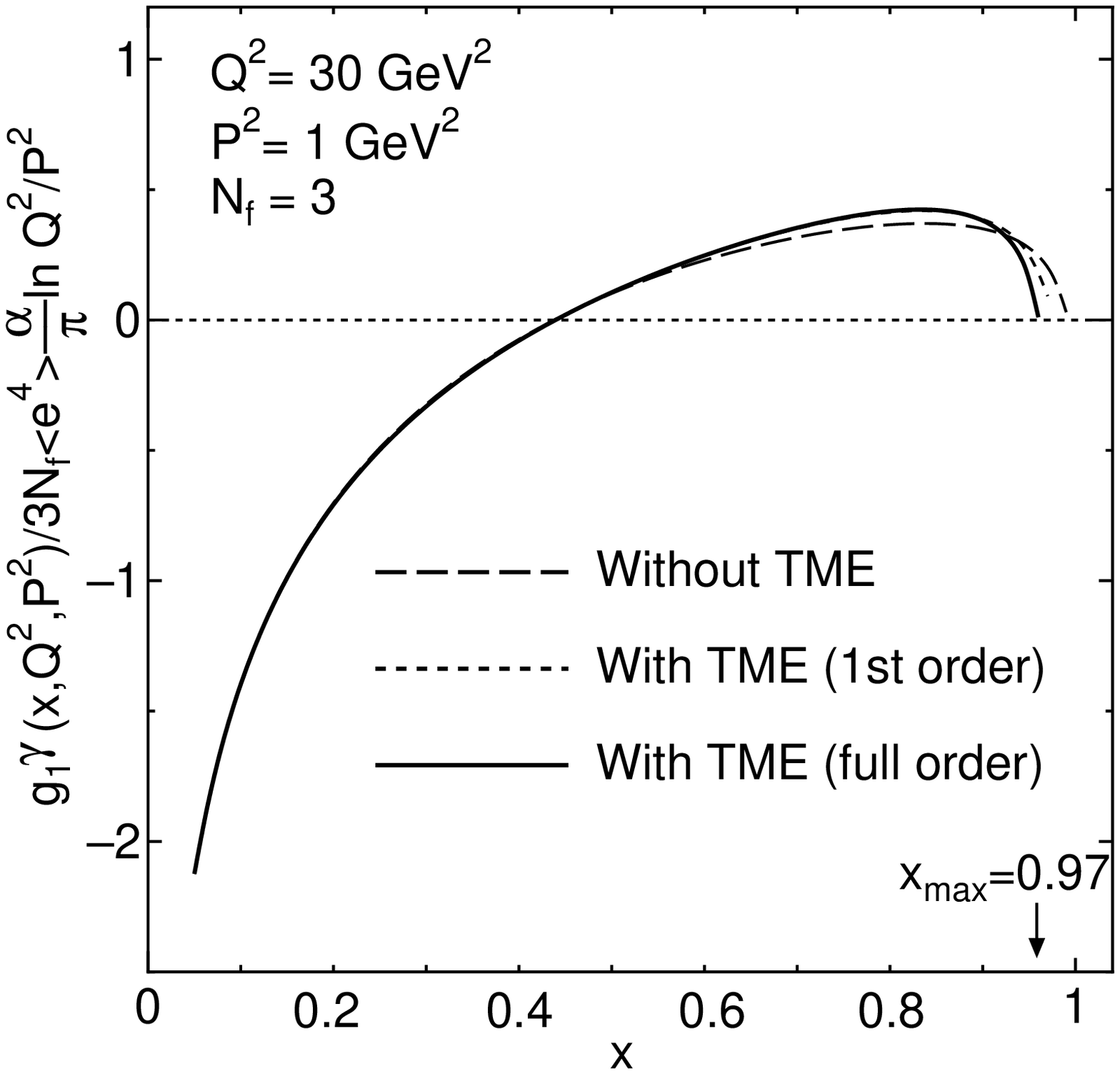,height=2.2in,width=2.2in}
\vspace*{-3.4in}
{\epsfig{file=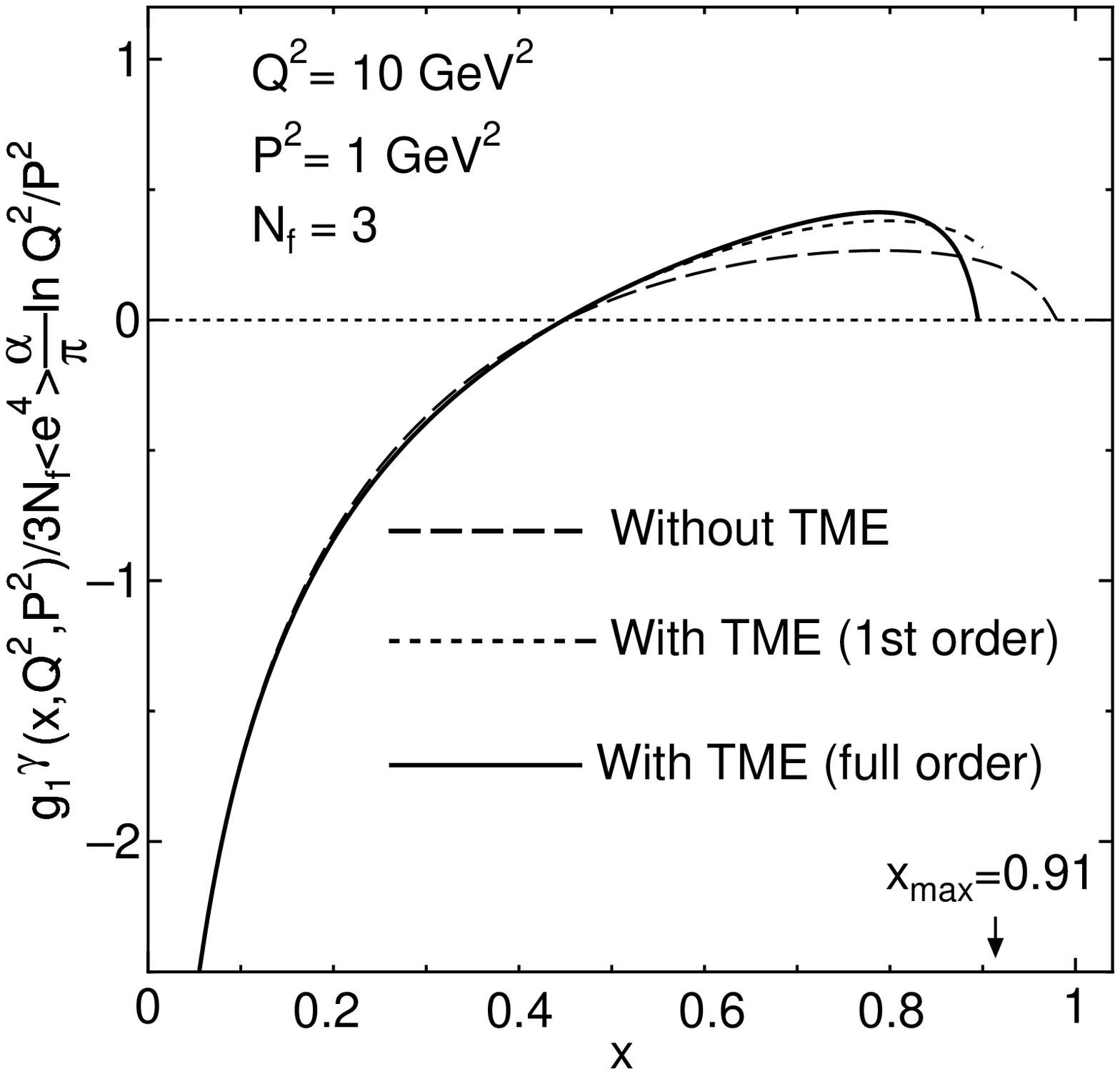,height=2.2in,width=2.2in}}
\vspace*{3.5in}
\caption{$g_1^\gamma(x,Q^2,P^2)$ with full TME (solid curve),
with the first order TME (short-dashed curve) and without TME 
(dashed curve), for $Q^2=30$ and $10$ GeV$^2$ with $P^2=1$ GeV$^2$.}
\label{fig3}
\end{figure}

\vspace{5cm}

In Figure 3, we have shown $g_1^\gamma(x,Q^2,P^2)$ with TME 
as a function of $x$ (solid curve) for
$Q^2=30$ GeV$^2$  (upper) with $P^2=1$ GeV$^2$ and 
for $Q^2=10$ GeV$^2$ (lower) 
with $P^2=1$ GeV$^2$. The vertical axis is in units of
$3N_f\langle e^4\rangle \frac{\alpha}{\pi}\ln(Q^2/P^2)$, where
$\alpha=e^2/4\pi$, the QED coupling constant,  
$N_f$ is the number of active flavors, 
and $\langle e^4\rangle =\sum_{i=1}^{N_f}e_i^4/N_f$ with 
$e_i$ being the electric charge of $i$th flavor quark.
Also plotted are  $g_1^\gamma(x,Q^2,P^2)$ without
TME (dashed curve) and the one with TME
included up to the first order in $P^2/Q^2$ (short-dashed curve).
We observe that the target mass effects appear between intermediate $x$ and 
$x_{\rm max}$, and that the effects become sizable 
when the ratio $P^2/Q^2$ is increased.
The distinction between the behaviors of $g_1^\gamma$ with
and without  TME is remarkable near $x_{\rm max}$. We get  
$x_{\rm max}\approx 0.97$ for $Q^2=30$ GeV$^2$ with 
$P^2=1$ GeV$^2$ and $x_{\rm max}\approx 0.91$ for $Q^2=10$ GeV$^2$ 
with $P^2=1$ GeV$^2$. The graphs of $g_1^\gamma$ with TME vanish at $x_{\rm max}$
as they should. 

\section{QCD SUM RULES WITH TME}

If
TME is not taken into account, 
the polarized virtual photon structure function $g_1^\gamma(x,Q^2,P^2)$ 
satisfies the following sum rule \cite{NSV93,SU}: 
\begin{eqnarray}
&&\hspace{-0.7cm}
\Gamma_1^\gamma\equiv\int^1_0 dx 
g_1^\gamma(x,Q^2,P^2)=-\frac{3\alpha}{\pi}
\sum_{i=1}^{N_f}e_i^4+{\cal O}(\alpha_s).\nonumber\\
\hspace{-0.8cm}\label{Firstg1Without}
\end{eqnarray}
Note for the real photon target we have the vanishing sum rule \cite{BBS98}.
The right-hand side corresponds to the twist-2 contribution,  
and actually the first term is the consequence of the QED axial anomaly. 
Now it will be interesting to see how this
is modified when TME is included.

Once the target mass corrections are taken into account, 
the above sum rule is modified to  
the first Nachtmann moment, which reads 
\begin{eqnarray}
&&\hspace{-0.5cm}\frac{1}{9}\int_0^{x_{\rm max}}dx
\frac{\xi^2}{x^2}\left[
5+4\sqrt{1-\frac{4P^2x^2}{Q^2}}\right]g_1^\gamma(x,Q^2,P^2)\nonumber\\
&&\hspace{-0.5cm}+\frac{4}{3}\int_0^{x_{\rm max}}dx
\frac{\xi^2}{x^2}\frac{P^2x^2}{Q^2}g_2^\gamma(x,Q^2,P^2)\nonumber\\
&&\hspace{-0.5cm} = -\frac{3\alpha}{\pi}\sum_{i=1}^{N_f}e_i^4+{\cal O}(\alpha_s)~.\label{Firstg1With}
\end{eqnarray}
The 1st moment of $g_1^\gamma$ with TME for the nucleon target was discussed
in refs. \cite{MU,KU}.
The power-series expansion in $P^2/Q^2$ gives the first order TME,
the difference of LHS's of (\ref{Firstg1With}) and (\ref{Firstg1Without}):
\begin{eqnarray}
&&\hspace{-0.5cm}\Delta\Gamma_1^\gamma
=-\left\{\frac{2}{9}M_2^{n=3}+\frac{8}{9}M_3^{n=3}\right\}
\frac{P^2}{Q^2}\nonumber\\
&&\hspace{3cm}+{\cal O}\left((P^2/Q^2)^2\right).
\end{eqnarray}

The 1st moment for $g_2^\gamma$ without TME known as
the Burkhardt-Cottingham sum rule \cite{BC}:
\begin{equation}
\int_0^1dx \ g_2^\gamma(x,Q^2,P^2)=0
\end{equation}
turns into the one with TME given by
\begin{equation}
\int_0^{x_{\rm max}}dx \ g_2^\gamma(x,Q^2,P^2)=0~,
\end{equation}
where the upper-limit of the integration has changed from 1
to $x_{\rm max}$.

\section{CONCLUSION}

To summarize we have studied the target mass effects in the 
virtual photon's spin structure functions, $g_1^\gamma(x,Q^2,P^2)$ 
and $g_2^\gamma (x,Q^2,P^2)$, which can be measured in the future 
experiments of the polarized version of the $ep$ or $e^+e^-$ colliders.
The evaluation of kinematical target mass effects is 
considered to be important to
extract dynamical higher-twist effects. 

We have derived the expressions for $g_1^\gamma (x,Q^2,P^2)$ 
and $g_2^\gamma (x,Q^2,P^2)$ in closed form by inverting the
Nachtmann moments for the twist-2 and twist-3 operators.
Our numerical analysis shows that the target mass effects 
appear at large $x$ and become sizable near 
$x_{\rm max}(=1/(1+\frac{P^2}{Q^2}))$, as the ratio $P^2/Q^2$ increases. 

Here we have also examined 
the target mass effects for the first-moment sum rules of $g_1^\gamma$ and 
$g_2^\gamma$. For the kinematic region we consider, the corrections to 
the first moment of $g_1^\gamma$ turn out to be negligibly small. The 
first moment of $g_2^\gamma$ leads to the Burkhardt-Cottingham sum rule, where
only change exists in the upper limit of integration from 1 to $x_{\rm max}$.

There still remain two important subjects to be studied. The first one is
the quark-mass effects in $g_1^\gamma$ and $g_2^\gamma$. The heavy flavor
contribution to $g_2(x,Q^2)$ for the nucleon 
has been explored in ref.\cite{BRN}. It would be intriguing to investigate 
the polarized photon case. Another subject yet to be
studied is the transition from real to virtual photon, especially the 
1st moment sum rule; how to reconcile the vanishing sum for the former
with the non-vanishing one for the latter.

\end{document}